# Resonance-type thickness dependence of optical second harmonic generation in thin-films of the topological insulator Bi$_2$Se$_3$

Yuri D. Glinka,[1,2]* Sercan Babakiray,[1] Trent A. Johnson,[1] Mikel B. Holcomb,[1] and David Lederman[1]

[1]*Department of Physics and Astronomy, West Virginia University, Morgantown, WV 26506-6315, USA*
[2]*Institute of Physics, National Academy of Sciences of Ukraine, Kiev 03028, Ukraine*

Optical second harmonic generation (SHG) has been measured for the first time in reflection from the nanometer-thick films (6 to 40 nm) of the topological insulator Bi$_2$Se$_3$ using 1.51 eV (820 nm) Ti:Sapphire laser photons and revealed a strong dependence of the integral SHG intensity on the film thickness. The integral SHG intensity was determined by area integration of the SHG rotational anisotropy patterns measured for different input-output light polarization geometries. A ~100-fold enhancement of the integral SHG intensity with decreasing film thickness has been suggested to result from the DC-electric-field-induced SHG (EFISHG) effects. Two sources of dynamically created DC electric field were proposed: (i) the capacitor-type DC electric field that gradually increases with decreasing film thickness from 40 to 6 nm due to a dynamical imbalance of photoexcited long-lived carriers between the opposite-surface Dirac surface states and (ii) an DC electric field associated with a nonlinearly excited Dirac plasmon, which is responsible for the resonant enhancement of the integral SHG intensity for the 10 nm thick film with a Lorentz-shaped resonance of ~1.6 nm full width at half maximum. Additionally to the general SHG enhancement trends with decreasing film thickness, a relative decrease of the out-of-plane contribution with respect to the in-plane contribution was observed. Using a theoretical treatment of the measured SHG rotational anisotropy patterns, this effect has been suggested to result from the joint contributions of the linear and quadratic DC electric field effects to the EFISHG response.

## I. INTRODUCTION

After the first nonlinear optics experiment carried out in 1962 by Franken *et al.*[1] showed that a 3-joule red ruby laser pulse focused onto a quartz crystal generates a few nanojoules of ultraviolet light at exactly twice the incident frequency [optical second harmonic generation (SHG)], the effect quickly received wide attention from the semiconductor community as well as from technology research agencies. This considerable interest in SHG has been driven by the prospect of using SHG as a non-destructive method of surface structural analysis if it is measured in reflection from semiconductor media. The basic idea for this SHG application was proposed by Bloembergen *et al.*[2] in 1968, who pointed out that the semiconductor surface structural symmetry can be monitored by the SHG process since it is governed by a tensor quantity that contains elements of the crystal symmetry. Consequently, the electric-dipole contribution has been suggested to dominate the SHG effect as a lowest-order multipole contribution to the bulk [three-dimensional (3D)] nonlinear polarization. Since a semiconductor's true surface can be modeled as an abrupt profile approximately one nanometer thick, the semiconductor surface [two-dimensional (2D)] symmetry is usually different from that of the 3D crystal symmetry. The SHG method is therefore an interface (surface) sensitive probe of the nanometer scale, especially if the bulk contribution is forbidden by the crystal symmetry. The spatial sensitivity of the reflected SHG response offers an advantage with respect to the usual (linear) optical reflectivity.

Somewhat later, Guidotti *et al.*[3] observed and comprehensively studied the orientation dependence of SHG in reflection from Si and Ge as a function of light polarization, crystal surface symmetry, and an angle of rotation of the surface about the surface normal. This experimental approach allowed them to monitor the independent lattice polarizations corresponding to the certain tensor components as a function of a specific surface symmetry. The resulting SHG rotational anisotropies, which are consistent with the Si and Ge cubic centrosymmetric crystal symmetry of different crystal faces, have also been considered theoretically by Sipe *et al.*,[4] thus providing a good theoretical foundation for the explanation of SHG experimental data. Similarly, SHG in reflection from non-centrosymmetric III-V semiconductors, such as GaAs, has been widely studied and vigorously debated in terms of the consistency of the surface rotational anisotropy and the surface symmetry of different crystal faces.[5-7]

One of the most intriguing effects regarding the SHG sensitivity to interfaces is the laser-induced phase transitions in semiconductors with either diamond (Si) or zinc-blende (GaAs) lattice structure, which has been the scope of intensive experimental and theoretical studies since the discovery of pulsed-laser annealing.[8-18] Because Si is a centrosymmetric semiconductor, the electric-dipole contribution to the SHG process is allowed by symmetry only at the surface of the crystal. Moreover, because of the surface crystalline rotational anisotropy, the observation of the SHG response is possible only through the careful choice of polarization and sample orientation in order to measure the non-zero elements of the nonlinear second-order optical susceptibility tensor [ $\chi^{(2)}$ ]. Alternatively, GaAs is a non-centrosymmetric material and therefore the electric-dipole contribution to SHG is symmetry allowed for both the bulk and the surface crystal structures. The observed drop of the SHG intensity, and hence the second-order nonlinearity, for GaAs was explained in terms of the hypothesis that the near-surface layer of the semiconductor undergoes a plasma-induced structural phase transition into a centrosymmetric state where SHG is symmetry forbidden in the electric-dipole approximation.

The spatial sensitivity of SHG makes this technique uniquely promising for studying Dirac surface states (SS) in the topological insulators (TIs) that have a 3D insulator-type bandgap (for Bi$_2$Se$_3$, for example, $E_g$ ~ 0.3 eV) and protected 2D gapless conducting phase on their surface due to the combination of strong spin-orbit interactions and time-reversal symmetry.[19,20] Recent studies of SHG measured in reflection from the single crystals of Bi$_2$Se$_3$(111) using ~1.56 eV (795 nm) Ti:Sapphire laser photons showed that owing to the centrosymmeric nature of Bi$_2$Se$_3$ crystals, the SHG rotational anisotropy is governed by a few atomic layers near the surface, the range where the 2D Dirac SS predominantly contribute to the carrier dynamics.[21-23] However, the finite-size effect on the SHG response from thin-films of the TI Bi$_2$Se$_3$ still remains unknown despite having more prospects on applications in 2D electronic devices. It should also be noted that the sensitivity of the SHG response to Dirac SS is, in principle, at least comparable to that of the angle-resolved photoemission spectroscopy (ARPES),[24-28] a technique whose surface sensitivity originates from the small penetration depth of more energetic incident light (a few nm) and the small escape depth of photoemitted electrons.

It has also been suggested that the SHG response from Bi$_2$Se$_3$(111) single crystals is not purely surface-related and hence is not exclusively governed by the second-order nonlinearity.[21-23] The 3D contribution induced by a depletion DC electric field appearing through the third-order nonlinearity [DC-electric-field-induced SHG (EFISHG)] must also to be taken into account.[29,30] This depletion-field-induced effect is known to predominantly contribute to the EFISHG response from non-centrosymmetric bulk semiconductors.[31] However, because the third-order and second-order susceptibilities for the single crystals of Bi$_2$Se$_3$(111) have the same symmetry constraints,[23] the depletion electric field effect has been found to only enhance the total intensity of SHG from the near-surface region without any changes in the rotational anisotropy patterns. The width of the surface depletion layer for Bi$_2$Se$_3$(111) single crystals varies in the range 2 - 40 nm depending on the free carrier density.[32] This short-range depletion layer can be a key factor in modifying the properties of 3D crystalline Bi$_2$Se$_3$ when switching to thin films of only a few nanometers thick.

Consequently, for Bi$_2$Se$_3$ thin films a crossover of the 3D TI to the 2D limit (gapped Dirac SS) has recently been observed if the thickness is below six quintuple layers (~ 6 nm).[33] This kind of finite-size effect has initially been suggested to result from the time-reversal symmetry breaking due to direct coupling between boundary modes from the opposite-surface Dirac SS. However, within the framework of depletion electric field concept, the Dirac SS on opposite surfaces of Bi$_2$Se$_3$ films can interact with each other even at longer distances than those required for direct intersurface coupling. Because of the band structure distortion near the surface due to space-charge accumulation,[32,34] which depletes 3D electrons across the film thickness below ~20 nm thick, the indirect intersurface coupling occurs due to the interaction between depletion electric fields when the sum of depletion layer widths associated with each surface of the film exceeds the film thickness.[35] The indirect intersurface coupling is expected to be weaker than the direct intersurface coupling and hence it does not affect the gapless Dirac SS whereas efficiently redistributes free carriers towards the surfaces with decreasing film thickness. The latter behavior has been suggested to result in an increase of 2D carrier density with decreasing film thickness and in the corresponding surface-dominated Hall conductivity.[32,36-38] Furthermore, the 3D-carrier-depletion-induced indirect intersurface coupling in films below ~20 nm thick has been suggested to be responsible for the metallic-type carrier relaxation rate and an enhancement of the recombination rate in 2D Dirac SS.[34,35] The indirect intersurface coupling between Dirac SS in thin films of the TI Bi$_2$Se$_3$ seems to be similar to that occurring between Dirac SS in electrostatically coupled graphene bilayers.[39-42] This kind of electrostatic coupling suggests the same type surface plasmon modes to be excited for both systems. Consequently, the quantum interference between surface phonon and Dirac plasmon states in Bi$_2$Se$_3$ films below ~20 nm thick has recently been observed using Raman spectroscopy,[36] in a similar way as that observed for graphene bilayers in infrared absorption.[43] These observations suggest that the indirect intersurface coupling effect in thin films of the TI Bi$_2$Se$_3$ should also be efficiently monitored through the EFISHG response.

It has also been found that the photoexcitation of thin films of the TI Bi$_2$Se$_3$ leads to the quasi-steady Fermi energy difference at the opposite-surface Dirac SS due to a dynamical imbalance in the photoexcited long-lived carrier densities.[35] This behavior gives rise to the development of the 3D-carrier-depletion-mediated capacitor-type electric field directed along the film normal. Because the dynamical charge imbalance in semiconductor multilayers is known to be an efficient source of the EFISHG response,[29,30,44-48] the SHG technique is expected to be successfully applied for studying photoexcited carrier dynamics in thin films of the TI Bi$_2$Se$_3$. Finally, a recent observation of the Dirac-plasmon-enhanced Raman responses from Bi$_2$Se$_3$ thin films[36] and theoretical predictions for the giant plasmon-enhanced SHG in graphene and semiconductor 2D electron systems[49] makes SHG technique also suitable for studying plasmon-related phenomena in TIs and 2D material stacks.

In this paper, we present a first experimental study on the stationary SHG rotational anisotropy patterns of Bi$_2$Se$_3$ thin films ranging in thickness from 6 to 40 nm. We showed that the SHG rotational anisotropy patterns measured in reflection from Bi$_2$Se$_3$ films using 1.51 eV (820 nm) Ti:Sapphire laser photons are almost identical to those measured for the single crystals of Bi$_2$Se$_3$(111). However, the integral intensity of the SHG responses, which was determined by area integration of the SHG rotational anisotropy patterns measured for different input-output light polarization geometries, and the relative intensity of the rotational anisotropy components are found to be strongly dependent on the film thickness. In particular, we observed a ~10-fold increase of the integral SHG intensity with decreasing film thickness from 40 to 6 nm. We proposed that this enhancement is due to the EFISHG contribution induced by the 3D-carrier-depletion-mediated capacitor-type electric field, the strength of which gradually increases with decreasing film thickness owing to the dynamical charge imbalance between the opposite-surface Dirac SS. We also observed another ~10-fold resonant enhancement of the integral EFISHG intensity for the 10 nm film with a Lorentz-shaped resonance of ~1.6 nm full width at half maximum (FWHM). This resonant feature has been suggested to originate from the DC electric field associated with a nonlinearly excited Dirac plasmon. We also observed a relative decrease of the out-of-plane contribution to the EFISHG response with respect to the in-plane contribution when the film thickness decreases from 40 to 6 nm. This effect is associated with the joint contributions of the linear and quadratic DC electric field effects to the EFISHG response.

It should be noted that other sources of the resonant enhancement of SHG from $Bi_2Se_3$ films seem to be less probable. Specifically, the strain-induced SHG enhancement, which usually manifests itself in thin films,[50,51] is expected to be weak for several reasons. First, the effect of the strain on the optical properties with decreasing $Bi_2Se_3$ film thickness deduced from Raman spectroscopy is ~1%.[36] Secondly, the strain-induced bandgap change in $Bi_2Se_3$ is known to be negligible for the film thickness range used.[52] Furthermore, the change of the lattice constants that we measured using x-ray diffraction can be estimated to be as small as <0.5% over the entire range of film thicknesses used. Additionally, the strain-induced enhancement of the SHG intensity usually shows a smooth dependence with increasing strain,[50,51] i.e. it disagrees with our resonance-type observation. In contrast, the observed ~2-fold increase of the absorption coefficient of $Bi_2Se_3$ films with decreasing film thickness from 40 to 6 nm has been evidenced to be due to the free carrier (Drude) absorption with increasing free carrier density.[34,36] The latter fact verifies that $Bi_2Se_3$ is an example of heavily doped semiconductors (~$10^{19}$ $cm^{-3}$), where the optical excitation we used (~$10^{19}$-$10^{20}$ $cm^{-3}$),[34] the photoexcited carrier redistribution dynamics between the bulk and Dirac SS as well as carrier collective motion (plasmons) significantly modify the optical properties of these materials ranging in thickness from 6 to 40 nm rather than the strain-induced effects. It is also important to note that the interference effects, usually occurring in optical nonabsorbing slabs where the incident fundamental and the reflected fundamental and SHG beams possibly follow multiple reflections within the slab,[53-55] cannot be responsible for the resonance-type thickness dependence observed as well, since the penetration depth of the fundamental and SHG light (10-25 nm) is comparable or less than the film thickness.[23,34] Moreover, because of the EFISHG process is expected to be dominant, the SHG escape depth is completely controlled by the depletion electric field width,[32,35] which for free carrier densities presenting in our films ranging in thickness from 40 to 6 nm (0.5 – 3.5×$10^{19}$ $cm^{-3}$)[36] can be estimated to be as long as 12 – 5.2 nm,[35] respectively. The latter estimates completely eliminate any interference effects to be considered. Finally, we note that because our films were grown on the sapphire substrates, which are transparent in the visible range, a novel optical behavior observed for highly absorbing thin semiconductor films grown on the metallic substrates also can be eliminated from consideration,[56] since similar situation for non-trivial interface phase shifts and the total phase accumulation can be realized for sapphire coated with thin semiconductor layers in the mid-infrared region, not in the visible region.

## II. EXPERIMENTAL DETAILS

The stationary SHG measurements for thin films of the TI $Bi_2Se_3$ were performed using the multifunctional pump-probe setup, which has been used in our previous studies of ultrafast carrier dynamics in these films.[34,35] The sketch of the experimental setup is shown in Fig. 1(a), where all functional capabilities have also been indicated explicitly. In this paper we present and discuss only experimental results on stationary SHG rotational anisotropy, which have been obtained exploiting exclusively the probe beam, whereas the normal incidence pump beam was used for the sample alignment procedure. The setup includes a Ti:Sapphire laser of the output average power of 2.5 W with pulse duration $\tau_L$= 100 fs, center photon energy 1.51 eV and repetition rate 80 MHz. A laser beam (probe) of various average powers in the range of 100 – 580 mW was at an incident angle of ~ 15° and focused to a spot diameter of ~ 100 μm. The SHG response measured in reflection geometry was collected by a photomultiplier tube (PMT) and the signal was measured with a lock-in amplifier triggered at the incident beam modulation frequency of 800 Hz. The SHG rotational anisotropy measurements were performed by rotating the sample mounted on a rotation stage about the surface normal with a step size of 2°. The sample was aligned to match the center of rotation with the center of the incident probe beam spot on the sample surface using the normal incidence pump beam, which afterwards was blocked for measurements. The linear polarization of incident fundamental and outgoing SHG beams was controlled to be either P (in the plane of incidence) or S (in the plane of the film) [Fig. 1(b)]. Four different light polarization geometries for the incident laser beam (in) and the outgoing SHG beam (out) were used for measuring rotational anisotropy patterns: $P_{in}$ - $P_{out}$, $S_{in}$ - $P_{out}$, $P_{in}$ - $S_{out}$, and $S_{in}$ - $S_{out}$. Experiments were performed in air and at room temperature. No film damage was observed for the laser powers used in the measurements reported here.

Experiments were performed on $Bi_2Se_3$ thin films that were 6, 8, 10, 12, 15, 20, 25, 30, 35, and 40 nm thick. The films were grown on 0.5 mm $Al_2O_3$(0001) substrates by molecular beam epitaxy, with a 10 nm thick $MgF_2$ capping layer to protect against oxidation. The growth process was similar to that previously described.[57] The polycrystalline $MgF_2$ capping layer was grown at room temperature without exposing the sample to atmosphere after the $Bi_2Se_3$ growth. The film thickness was determined from x-ray reflectivity (XRR) measurements. Reflection high-energy electron diffraction and x-ray diffraction showed that the $Bi_2Se_3$ films were epitaxial in the plane and highly crystalline out of the plane of the film. This finding is consistent with Raman measurements of the films,[36] where the expected phonon modes for $Bi_2Se_3$ were observed. Transport measurements revealed that all films have *n*-doping level in the range of 0.5 – $3.5 \times 10^{19}$ cm$^{-3}$,[36] which is also typical for as-grown $Bi_2Se_3$ films and single crystals.

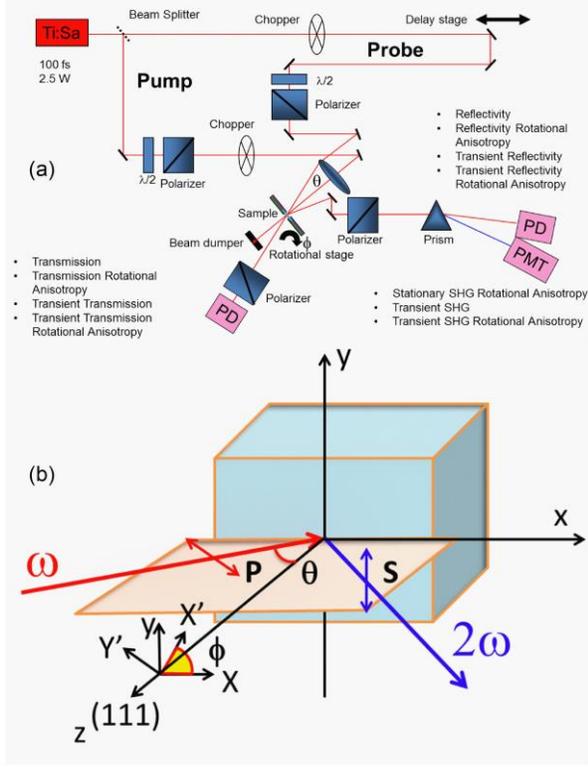

FIG. 1. (Color online) (a) Multifunctional experimental setup for optical study of thin films of the TI $Bi_2Se_3$. (b) The sketch of the experimental geometry and the light polarizations used for SHG rotational anisotropy measurements (see all notations in the text).

### III. EXPERIMENTAL RESULTS AND DISCUSSION

Figure 2(c) shows the SHG rotational anisotropy patterns for the 10 nm film measured at different light polarization geometries mentioned in the preceding section. It should be noted first that owing to the presence of the $MgF_2$ capping layer, in contrast to the freshly cleaved single crystals of $Bi_2Se_3$(111),[21-23] our films showed the highly stable SHG responses, suggesting the absence of any radiation-induced degradation and contaminant-related modification of the film surfaces. We note that because we did not observe any long-time variations in the SHG intensity, the effect of the interfacial DC electric fields, which can possibly arise from the trapping dynamics of photoexcited carriers in $MgF_2$ and $Al_2O_3$,[44] is assumed to be negligibly small. Nevertheless, the rotational anisotropy patterns revealed features similar to those observed for the freshly cleaved single crystals of $Bi_2Se_3$(111). This circumstance suggests the same rotational symmetry for the (111) surface of our films, where the crystal symmetry is reduced from $D_{3d}^5$ to $C_{3v}$.[21-23] The reduction in symmetry implies that the electric-dipole contribution to SHG from the bulk is forbidden by inversion symmetry due to the centrosymmetric nature of $Bi_2Se_3$ and therefore inversion symmetry breaking at the surface is assumed to govern exclusively the surface-related SHG response.

This behavior can be explained by assuming that the induced effective second order nonlinear polarization includes both surface and bulk contributions,[58]

$$\mathbf{P}_{eff}(2\omega) = \mathbf{P}_S F^{(S)}(\omega) F^{(S)}(2\omega) + \mathbf{P}_B F^{(B)}(\omega) F^{(B)}(2\omega) L_{eff}, \quad (1)$$

where $\mathbf{P}_S$ is the surface nonlinear polarization and $\mathbf{P}_B$ is the bulk nonlinear polarization; $F(\omega)$ and $F(2\omega)$ are the Fresnel factors for the incident input and output fields for the surface and bulk contributions, $L_{eff} = (k_{\omega,z} + k_{2\omega,z})^{-1} \approx 2(\alpha_{\omega,z} + \alpha_{2\omega,z})^{-1}$ = 14.3 nm is the effective phase-matching distance in $Bi_2Se_3$ for the incident fundamental and outgoing SHG beam, where $k$'s are $z$ components of the corresponding wave vectors and $\alpha_{\omega,z} = 0.04$ nm$^{-1}$ and $\alpha_{2\omega,z} = 0.1$ nm$^{-1}$ are the corresponding absorption coefficients.[23,34] The latter estimate exceeds the depletion layer widths associated with each surface of the film[35] and hence suggests that the experimental conditions are suitable for studying EFISHG in thin-films of the TI $Bi_2Se_3$. Subsequently, the bulk nonlinear polarization can be generalized as[59]

$$\mathbf{P}_B(2\omega) = \mathbf{P}_B^{(2)}(2\omega) - \nabla \cdot \mathbf{Q}_B^{(2)}(2\omega) + (c/2i\omega)\nabla \times \mathbf{M}_B^{(2)}(2\omega), \quad (2)$$

where $\mathbf{P}^{(2)}$, $\mathbf{Q}^{(2)}$, and $\mathbf{M}^{(2)}$ represent the electric-dipole, electric-quadrupole, and magnetic-dipole polarizations, respectively. Since higher-order multipole contributions are expected to be much weaker than the electric-dipole contribution, we neglect them in our consideration further below. This approximation is common for studying SHG in reflection from the anisotropic non-magnetic media of bulk crystalline centrosymmetric and non-centrosymmetric materials.[59,60] This simplification yields

$$\mathbf{P}_B(2\omega) = \chi_{ijk}^{(2)} E_i(\omega) E_j(\omega), \quad (3)$$

where $\chi_{ijk}^{(2)}$ is the second-order susceptibility tensor of $Bi_2Se_3$, $E_i(\omega)$ and $E_j(\omega)$ denote the driving electric fields of the incident light and $i, j, k$, are the induced polarization indices. Due to inversion symmetry of the centrosymmetric $Bi_2Se_3$ crystalline lattice, $\chi_{ijk}^{(2)}$ vanishes in the electric-dipole approximation and the effective nonlinear polarization of Eq. (1) is completely determined by the surface contribution, which can be expressed similarly to that of Eq. (3) but with the second-order susceptibility tensor of the surface $\chi_{ijk}^{(S)}$ replacing that of the bulk, such that

$$\mathbf{P}_{eff}(2\omega) = \chi_{ijk}^{(S)} E_i(\omega) E_j(\omega) F^{(S)}(\omega) F^{(S)}(2\omega). \quad (4)$$

The SHG intensity is then determined by the product of $\mathbf{P}_{eff}(2\omega)$ and its conjugate,[29,30,44-48]

$$I(2\omega) = A \left| \chi_{ijk}^{(S)} F_{ijk}^{(S)}(\omega) F_{ijk}^{(S)}(2\omega) \right|^2 I^2(\omega), \quad (5)$$

where $A$ is a constant and $I(\omega) = E_i(\omega) E_j(\omega)$ denotes the intensity of the incident laser light. The characteristic quadratic laser power dependence of the SHG response intensity can be verified experimentally. Figure 3 shows the power dependences of the integral SHG intensity (the area integration of the SHG rotational anisotropy patterns) measured for different light polarization geometries and for $Bi_2Se_3$ thin films of different thicknesses. The exact quadratic dependence obtained and the nonappearance of any rotational anisotropy changes with increasing laser power also indicate the absence of radiation-induced degradation of the films in the applied range of laser powers. It should be noted that this behavior is cardinally distinct from that observed in a lightly doped $Si(001)/SiO_2$ system, where the variation of rotational anisotropy with laser power has been suggested to be caused by carrier-induced screening of the DC electric field at the $Si(001)$–$SiO_2$ interface since the photoexcited carrier density was much higher than the doping level of Si wafer.[61] We associate the absence of the laser power effect on the rotational anisotropy of the SHG response from $Bi_2Se_3$ films with the aforementioned high doping level, which was comparable with the density of photoexcited carriers.

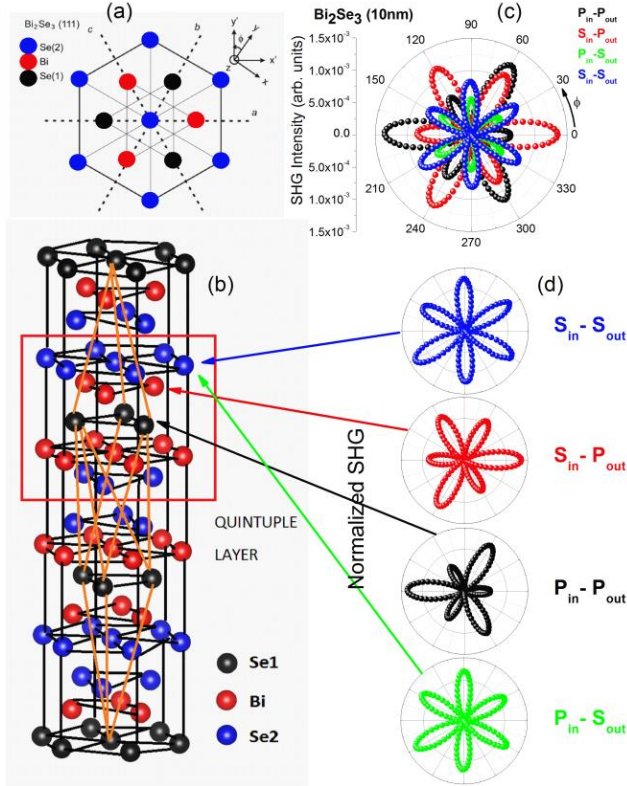
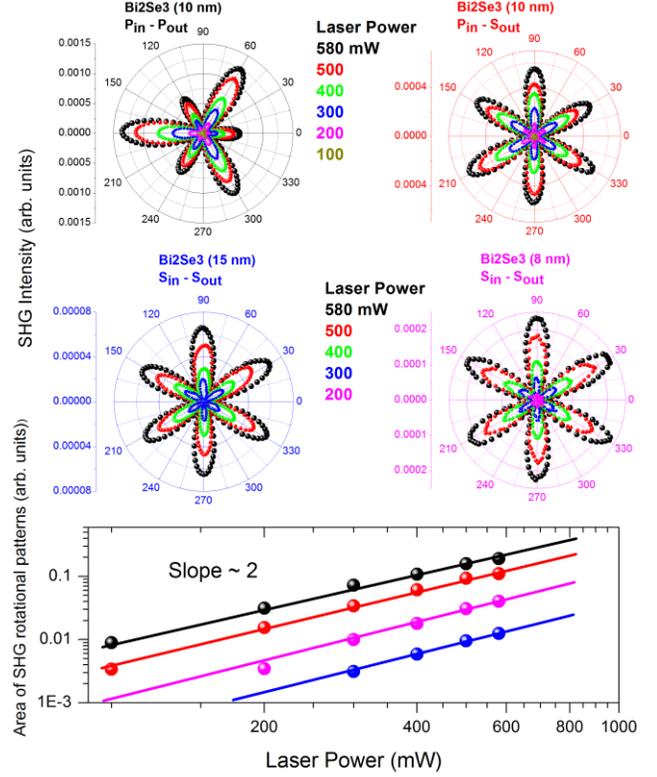

FIG. 2. (Color online) (a) Top view of the crystal unit cell structure of $Bi_2Se_3$ along the $z$-direction. $x$ and $y$ axis at the right top corner demonstrate the coordinate transformation at the sample rotation. (b) Rhombohedral unit cell of $Bi_2Se_3$ with three primitive lattice vectors (shown in orange). The red square indicates a quintuple layer with Se2-Bi-Se1-Bi-Se2 atoms, where (1) and (2) refer to different Se-atom lattice positions. (c) SHG rotational anisotropy patterns measured for the 10 nm thick $Bi_2Se_3$ film at different light polarization geometries indicated by the corresponding colors. (d) The same SHG rotational anisotropy patterns shown in (c) but being normalized and presented individually. The color arrows show the corresponding atomic layer assignment.

FIG. 3. (Color online) SHG rotational anisotropy patterns taken at different laser powers indicated by the corresponding colors are shown for several $Bi_2Se_3$ film thicknesses and light polarization geometries. Power dependences of the integral SHG intensity (the area of SHG rotational anisotropy patterns) are shown below as dots of the corresponding colors of the rotation coordinate frame. The straight solid lines present the quadratic function best fit to the data in log-log scale.

Depending on the selected output light polarization, the SHG rotational anisotropy patterns exhibit either sixfold or threefold rotational symmetry [Fig. 2(d)] in full agreement with symmetric arrangement of three upper atomic layers of $Bi_2Se_3$, for which the inversion symmetry breaking condition is most predominant [Figs. 2(a) and (b)]. The sixfold rotational symmetry measured in the $P_{in}$ - $S_{out}$, and $S_{in}$ - $S_{out}$ light polarization geometries with six equally-intense peaks separated by 60° and with no isotropic offset above the background indicates the predominant monocomponent nature of the SHG response. Alternatively, the threefold rotational symmetry measured in the $P_{in}$ - $P_{out}$, and $S_{in}$ - $P_{out}$ light polarization geometries with the three weak and three strong peaks separated by 120° indicates the multicomponent nature of the outgoing SHG response, i.e. the threefold rotational symmetry is not exact as a consequence of the interplay between multiple contributions to the SHG process.

The difference in SHG rotational symmetry between $S_{out}$ and $P_{out}$ light polarizations can be explained as follows. Because of the rhombohedral 3D symmetry of $Bi_2Se_3$, S-polarized and P-polarized incident beams induce both in-plane ($xy$) and out-of-plane (along the $z$-axis) nonlinear polarizations. However, the SHG response measured in outgoing S-polarization has a monocomponent nature and originates exclusively from the hyperpolarizability of the planar-hexagon-arranged Se-Se and Bi-Bi bonds, which have a sixfold symmetric arrangement [Fig. 2(a) and (b)].[21-23] Figure

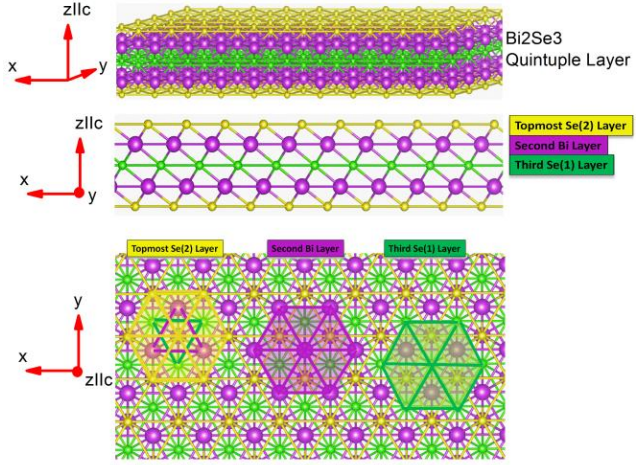

FIG. 4. (Color online) Crystal structure of $Bi_2Se_3$ quintuple layer visualized using VESTA software.[62] The rhombohedral crystal structure of $Bi_2Se_3$ presents the hexagonal planes of Se and Bi atoms stacked on top of each other along the $z$ direction. The top view (the lowest panel) indicates the hexagonal planes of Se and Bi atoms of three upper atomic layers by the corresponding color hexagons.

4 shows a visualization of the crystal structure of the $Bi_2Se_3$ quintuple layer using VESTA software,[62] which clearly demonstrate the existence of the global hexagonal continuous network of Se and Bi atomic planes stacked on top of each other along the $z$ direction. We note that our recent Raman scattering measurements allowed us to conclude that the topmost atomic layer in the films consists of hexagonally arranged Se-Se bonds (Fig. 4)[36] and hence the main contribution to the outgoing SHG response measured in S-polarization is governed by the hyperpolarizability of a continuous network of hexagonally arranged Dirac SS. Alternatively, the SHG response measured in outgoing P-polarization has a multicomponent nature and originates from both in-plane and out-of-plane nonlinear polarizations associated with the hyperpolarizability of the planar hexagon-arranged Se-Se and Bi-Bi bonds as well as Bi-Se bonds arranged into the rhombohedral unit cell along the trigonal axis (threefold rotational axis), respectively. The superposition of these two contributions of outgoing SHG determines the resulting shape of rotational anisotropy patterns measured in P-polarization. The assignment to particular atomic bonds seems to be reasonable because the optical (electric) nonlinear susceptibilities (polarizability per unit cell) in the macroscopic crystalline structures are related to hyperpolarizabilities that are microscopic quantities referring to a unit cell in the case of polymers, thin crystalline slabs, and single molecules.[63] This well-known relation between susceptibilities and hyperpolarizabilities is the reason why the nonlinear polarization of the rhombohedral crystal structure of $Bi_2Se_3$ can be associated with a set of the hexagonal planes of Se and Bi atoms stacked on top of each other along the z direction but shifted in the $xy$ plane as shown in Fig. 4. Correspondingly, because of the same tensor component indices in the nonlinear susceptibilities and in the corresponding hyperpolarizabilities,[63] the in-plane nonlinear susceptibility of the $Bi_2Se_3$ crystalline structure in thin films can be associated with the monolayer hyperpolarizability of hexagonally arranged Se and Bi atomic networks. On the contrary, the out-of-plane nonlinear susceptibility can be associated with the hyperpolarizability of Se-Bi bonds, which have a threefold arrangement in a unit cell since connecting the neighboring hexagonal atomic monolayers shifted with respect to each other. Because the three upper atomic layers unambiguously define the symmetry of the rhombohedral unit cell of $Bi_2Se_3$, the SHG rotational anisotropy patterns exactly depict all of the crystal unit cell symmetry constraints [Fig. 2(a) and (c)].

Despite this dual characterization of nonlinear polarizations, we will use further below the corresponding tensor components of the nonlinear susceptibilities, thus trying to treat the finite-size effect in our films with respect to the single crystals of $Bi_2Se_3(111)$. The 3D $Bi_2Se_3$ is characterized by a rhombohedral crystal structure with space group $D_{3d}^5 (R\bar{3}m)$, for which the representation formed by allowing three symmetry operations to act [Fig. 2(a)]: (i) rotation about (111) axis by $0^0$, $120^0$, and $240^0$; (ii) mirror reflections about planes $a$, $b$, and $c$; and (iii) inversion symmetry. Because the second-order susceptibility tensor [$\chi_{ijk}^{(S)}$] must obey the same symmetry conditions as the crystal, all of its 27 components become zero by applying the aforementioned symmetry operations. Once the inversion symmetry is broken at the surface, the crystal symmetry is reduced from $D_{3d}^5$ to $C_{3v}$ and, as a consequence, $\chi_{ijk}^{(S)}$ is given by[21,23]

$$\chi_{ijk}^{(S)} = \begin{pmatrix} \begin{pmatrix} \chi_{xxx} \\ 0 \\ \chi_{xxz} \end{pmatrix} & \begin{pmatrix} 0 \\ -\chi_{xxx} \\ 0 \end{pmatrix} & \begin{pmatrix} \chi_{xxz} \\ 0 \\ 0 \end{pmatrix} \\ \begin{pmatrix} 0 \\ -\chi_{xxx} \\ 0 \end{pmatrix} & \begin{pmatrix} -\chi_{xxx} \\ 0 \\ \chi_{xxz} \end{pmatrix} & \begin{pmatrix} 0 \\ \chi_{xxz} \\ 0 \end{pmatrix} \\ \begin{pmatrix} \chi_{zxx} \\ 0 \\ 0 \end{pmatrix} & \begin{pmatrix} 0 \\ \chi_{zxx} \\ 0 \end{pmatrix} & \begin{pmatrix} 0 \\ 0 \\ \chi_{zzz} \end{pmatrix} \end{pmatrix}, \quad (6)$$

which contains only four non-vanishing components of the form $\chi_{xxx} = -\chi_{xyy} = -\chi_{yxy}$, $\chi_{zxx} = \chi_{zyy}$, $\chi_{xxz} = \chi_{yyz}$, and $\chi_{zzz}$. In order to obtain the experimentally measured quantities, the sample coordinate frame is required to be transformed to the laboratory coordinate frame by using the standard 3×3 rotation matrix [Fig. 1(b)],

$$R(\phi) = \begin{pmatrix} \cos(\phi) & -\sin(\phi) & 0 \\ \sin(\phi) & \cos(\phi) & 0 \\ 0 & 0 & 1 \end{pmatrix}. \quad (7)$$

The application of this procedure modifies Eq. (6) to[21,23]

$$\chi_{ijk}^{(S)} = \begin{pmatrix} \begin{pmatrix} \chi_{xxx}\cos(3\phi) \\ -\chi_{xxx}\sin(3\phi) \\ \chi_{xxz} \end{pmatrix} & \begin{pmatrix} -\chi_{xxx}\sin(3\phi) \\ -\chi_{xxx}\cos(3\phi) \\ 0 \end{pmatrix} & \begin{pmatrix} \chi_{xxz} \\ 0 \\ 0 \end{pmatrix} \\ \begin{pmatrix} -\chi_{xxx}\sin(3\phi) \\ -\chi_{xxx}\cos(3\phi) \\ 0 \end{pmatrix} & \begin{pmatrix} -\chi_{xxx}\cos(3\phi) \\ \chi_{xxx}\sin(3\phi) \\ \chi_{xxz} \end{pmatrix} & \begin{pmatrix} 0 \\ \chi_{xxz} \\ 0 \end{pmatrix} \\ \begin{pmatrix} \chi_{zxx} \\ 0 \\ 0 \end{pmatrix} & \begin{pmatrix} 0 \\ \chi_{zxx} \\ 0 \end{pmatrix} & \begin{pmatrix} 0 \\ 0 \\ \chi_{zzz} \end{pmatrix} \end{pmatrix}. \quad (8)$$

Equation (8) allows for the assignment of different linear light polarization geometries that correspond to incident fundamental and outgoing SHG light as

$$I_{SS}(2\omega) = B\left|a^{(1)}\cos(\theta)\sin(3\phi)\right|^2, \quad (9)$$

$$I_{PS}(2\omega) = B\left|a^{(1)}\cos(\theta)\sin(3\phi)\right|^2, \quad (10)$$

$$I_{PP}(2\omega) = B\left|(a^{(3)}+a^{(2)})\sin(\theta) - a^{(1)}\cos(\theta)\cos(3\phi)\right|^2, \quad (11)$$

$$I_{SP}(2\omega) = B\left|a^{(2)}\sin(\theta) + a^{(1)}\cos(\theta)\cos(3\phi)\right|^2, \quad (12)$$

where the first and second sub-indices of $I(2\omega)$ denote the input and output light polarizations, respectively, $B = AI^2(\omega)$ is a constant for the given experimental conditions, $\phi$ is the angle between the scattering plane and the mirror ($xz$) plane of the crystal surface, and $\theta = 15^\circ$ is an angle of the laser light incidence [Fig. 1(b)]. Taking into account Eq. (5), one can conclude that the coefficients of Eqs. (9)-(12)

$$\begin{aligned} a^{(1)} &= \chi_{xxx} F_{xxx}^{(S)}(\omega) F_{xxx}^{(S)}(2\omega), \\ a^{(2)} &= \chi_{zxx} F_{zxx}^{(S)}(\omega) F_{zxx}^{(S)}(2\omega), \\ a^{(3)} &= 2\chi_{xxz} F_{xxz}^{(S)}(\omega) F_{xxz}^{(S)}(2\omega) + \chi_{zzz} F_{zzz}^{(S)}(\omega) F_{zzz}^{(S)}(2\omega), \end{aligned} \quad (13)$$

determine the amplitudes of the in-plane [$a^{(1)}$] and out-of-plane [$a^{(2)}$ and $a^{(3)}$] surface SHG responses.

Equations (9) and (10) together with the corresponding coefficients in the form of Eq. (13) prove that the in-plane SHG response has a monocomponent nature, which originates from the hyperpolarizability of a continuous network of hexagonally arranged Dirac SS and therefore is responsible for the sixfold SHG rotational anisotropy measured in the $P_{in}$ - $S_{out}$, and $S_{in}$ - $S_{out}$ light polarization geometries (Fig. 2). Alternatively, Eqs. (11) and (12) together with the corresponding coefficients in the form of Eq. (13) demonstrate that the SHG responses measured in the $P_{in}$ - $P_{out}$, and $S_{in}$ - $P_{out}$ light polarization geometries have a multicomponent nature and originates from both the hyperpolarizability of a continuous network of hexagonally arranged Dirac SS as well as Bi-Se bonds arranged into the rhombohedral unit cell along the trigonal axis. Consequently, the SHG rotational anisotropy patterns measured in the $P_{in}$ - $P_{out}$, and $S_{in}$ - $P_{out}$ light polarization geometries can be either sixfold or threefold, depending on the corresponding weight coefficients, which become more evident by expanding Eqs. (11) and (12), such that Eqs. (9) - (12) can be simplified to

$$I_{SS}(2\omega) = B\left|c_3 \sin^2(3\phi)\right|, \quad (14)$$

$$I_{PS}(2\omega) = B\left|c_3 \sin^2(3\phi)\right|, \quad (15)$$

$$I_{PP}(2\omega) = B\left|c_1 - c_2\cos(3\phi) + c_3\cos^2(3\phi)\right|, \quad (16)$$

$$I_{SP}(2\omega) = B\left|d_1 + d_2\cos(3\phi) + d_3\cos^2(3\phi)\right|, \quad (17)$$

where the weight coefficients $c_1 = \left|(a^{(3)}+a^{(2)})\sin(\theta)\right|^2$ and $d_1 = \left|a^{(2)}\sin(\theta)\right|^2$, $c_2 = \left|(a^{(3)}+a^{(2)})a^{(1)}\sin(2\theta)\right|$ and $d_2 = \left|a^{(2)}a^{(1)}\sin(2\theta)\right|$, and $c_3 = d_3 = \left|a^{(1)}\cos(\theta)\right|^2$ set the partial contributions of the isotropic, threefold, and sixfold rotational anisotropy components to the SHG responses, respectively.

We note that according to Eqs. (14) - (17), the sixfold rotational anisotropy component of the out-of-plane SHG response measured in the $P_{in}$ - $P_{out}$, and $S_{in}$ - $P_{out}$ light polarization geometries should be rotated by $30^\circ$ with respect to the sixfold rotational anisotropy component of the in-plane SHG response measured in the $S_{in}$ - $S_{out}$, and $P_{in}$ - $S_{out}$ light polarization geometries as a consequence of the cofunction identity $\cos^2(3\phi) = \sin^2(3(\phi+30^0))$. Additionally, the out-of-plane SHG response is expected to have a threefold rotational anisotropy component [second term in Eqs. (16) and (17)] and an isotropic component [first term in Eqs. (16) and (17)]. Because the threefold rotational anisotropy components in Eqs. (16) and (17) contribute to the SHG response measured in the $P_{in}$ - $P_{out}$, and $S_{in}$ - $P_{out}$ light polarization geometries with opposite signs, the corresponding rotational anisotropy patterns should be antisymmetric ($180^\circ$ rotation). The latter behavior is demonstrated in Fig. 2(d).

Now we turn to the more complicated case, nevertheless more consistent with reality. Because of the band structure bending near the surface due to space-charge accumulation,[32,35] which depletes 3D electrons near the surface, the $z$-directed depletion DC electric field ($\varepsilon_{DC}$) breaks 3D inversion symmetry by extending the $C_{3V}$ surface symmetry inward the film and therefore gives rise to the third-order process.[29,30,44-48] The term of DC electric field is applied as a consequence of the slowly varying electric field of frequency much lower than that of the driving electric field of the incident light wave (several hundred THz). Therefore, the electric field varying with frequency of several tens of THz or lower is actually the DC electric field. Consequently, the depletion electric field screening dynamics induced by photoexcited carrier population should be observable,[61] although probably in the time-resolved SHG responses. As we mentioned in the Introduction, owing to the same symmetry constraints for the third-order and second-order susceptibilities for $Bi_2Se_3(111)$ surface,[23] the depletion electric field only enhances the SHG response through the EFISHG effect without any changes in the shape of the rotational anisotropy patterns. Because the width of the surface depletion layer is comparable with the thickness range of $Bi_2Se_3$ films used in the current study,[35] the EFISHG effect is expected to be significant.

Subsequently, Eq. (3) can be modified to,[23,29,30,44-48]

$$\mathbf{P}_B(2\omega) = \left[\chi_{ijk}^{(2)} + \chi_{ijkl}^{(3)}\varepsilon_{DC}\right] E_i(\omega) E_j(\omega), \quad (18)$$

where $\chi^{(3)}_{ijkl}$ is the third-order susceptibility tensor. The induced effective nonlinear polarization of Eq. (1) then can be rewritten as

$$\mathbf{P}_{eff}(2\omega) = \begin{bmatrix} \chi^{(S)}_{ijk} F^{(S)}_{ijk}(\omega) F^{(S)}_{ijk}(2\omega) \\ + \chi^{(3)}_{ijkl} \mathcal{E}_{DC} F^{(B)}_{ijkl}(\omega) F^{(B)}_{ijkl}(2\omega) L_{eff} \end{bmatrix} \times E_i(\omega) E_j(\omega). \quad (19)$$

Following the same procedure as that for the depletion-field-independent surface and taking into account that $\chi^{(3)}$ has the same symmetry constraints as $\chi^{(2)}$, the SHG responses of Eqs. (14) – (17) can be rewritten in the same form but of modified weight coefficients,

$$c_1 = \left| \left( a^{(3)} + a^{(2)} + b^{(3)} + b^{(2)} \right) \sin(\theta) \right|^2,$$
$$d_1 = \left| \left( a^{(2)} + b^{(2)} \right) \sin(\theta) \right|^2,$$
$$c_2 = \left| \left( a^{(3)} + a^{(2)} + b^{(3)} + b^{(2)} \right) \left( a^{(1)} + b^{(1)} \right) \sin(2\theta) \right|, \quad (20)$$
$$d_2 = \left| \left( a^{(2)} + b^{(2)} \right) \left( a^{(1)} + b^{(1)} \right) \sin(2\theta) \right|,$$
$$c_3 = d_3 = \left| \left( a^{(1)} + b^{(1)} \right) \cos(\theta) \right|^2,$$

which now become DC-electric-field-dependent since

$$b^{(1)} = -\chi_{xyyz} F^{(B)}_{xyyz}(\omega) F^{(B)}_{xyyz}(2\omega) L_{eff} \mathcal{E}_{DC},$$
$$b^{(2)} = \chi_{zxxz} F^{(B)}_{zxxz}(\omega) F^{(B)}_{zxxz}(2\omega) L_{eff} \mathcal{E}_{DC}, \quad (21)$$
$$b^{(3)} = \begin{bmatrix} \chi_{xzxz} F^{(B)}_{xzxz}(\omega) F^{(B)}_{xzxz}(2\omega) + \chi_{xxzz} F^{(B)}_{xxzz}(\omega) F^{(B)}_{xxzz}(2\omega) \\ + \chi_{zzzz} F^{(B)}_{zzzz}(\omega) F^{(B)}_{zzzz}(2\omega) \end{bmatrix} L_{eff} \mathcal{E}_{DC},$$

where $\mathcal{E}_{DC}$ is directed along the z-axis.

Contracting DC electric field projections with tensor components, the weight coefficients of Eq. (21) can be simplified to the form similar to that for the depletion-field-independent (surface) contributions [Eq. (13)] but containing effective susceptibilities,[31]

$$b^{(1)} = -\chi^{(eff)}_{xyy} F^{(B)}_{xyy}(\omega) F^{(B)}_{xyy}(2\omega) L_{eff},$$
$$b^{(2)} = \chi^{(eff)}_{zxx} F^{(B)}_{zxx}(\omega) F^{(B)}_{zxx}(2\omega) L_{eff}, \quad (22)$$
$$b^{(3)} = \left[ 2 \chi^{(eff)}_{xxz} F^{(B)}_{xxz}(\omega) F^{(B)}_{xxz}(2\omega) + \chi^{(eff)}_{zzz} F^{(B)}_{zzz}(\omega) F^{(B)}_{zzz}(2\omega) \right] L_{eff}.$$

As it follows from this consideration, the depletion DC electric field contributes to the EFISHG response in a complicated way. The cross terms presenting in the weight coefficients of the form of Eq. (20) imply that the measured SHG responses may depend on $\mathcal{E}_{DC}$ both linearly and quadratically. Consequently, the shape of the rotational anisotropy patterns is expected to vary since the strength of the depletion DC electric field is known to increase with decreasing film thickness.[35] This tendency is demonstrated in Fig. 5, where the SHG rotational anisotropy patterns measured as a function of film thickness are shown. First we note that the thinnest film of 6 nm thick shows a distortion of the rotational symmetry, which is continuously repeatable for several films grown at the same conditions and hence indicates a thickness-related effect associated with a possible strained crystal structure involved. Secondly, despite the SHG rotational anisotropy patterns for $Bi_2Se_3$ thin films are similar to those measured for the single crystals of $Bi_2Se_3(111)$,[21-23] the relative intensity of the SHG peaks measured in outgoing P-polarization allows for significant variations with decreasing film thickness. Subsequently, relative intensities of the two sets of threefold rotational anisotropy peaks separated by 120° becomes more comparable with decreasing film thickness. This behavior suggests that the threefold rotational anisotropy component measured in the $P_{in}$ - $P_{out}$ and $S_{in}$ - $P_{out}$ light polarization geometries gradually decreases with respect to the sixfold rotational anisotropy component that predominantly defines the resulting shape of the SHG rotational anisotropy patterns with decreasing film thickness.

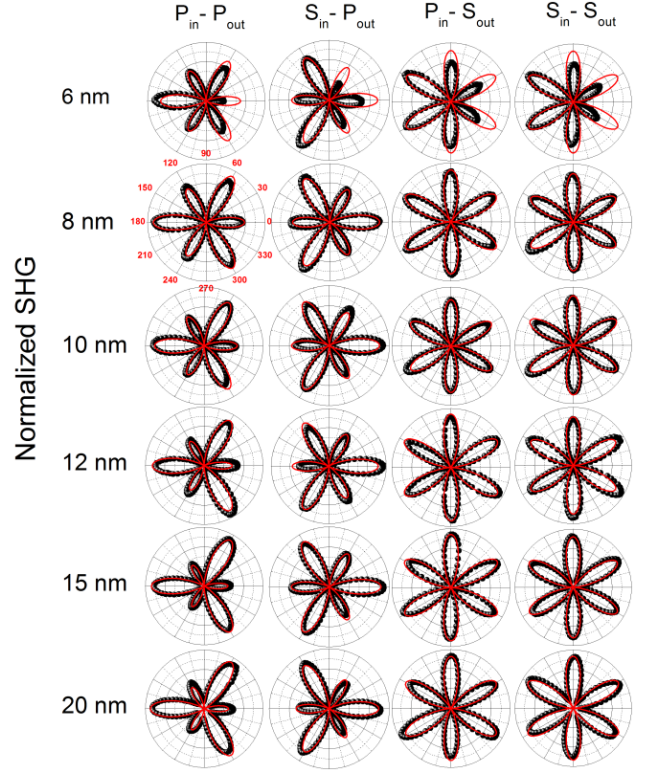

Fig. 5. (Color online) Normalized SHG rotational anisotropy patterns measured at different light polarization geometries (columns) for different $Bi_2Se_3$ film thicknesses indicated (rows). The red curves present best fits to the data using Eqs. (14) – (17) with the corresponding weight coefficients of the form of Eq. (20).

In general, the analysis of the finite-size effect on the SHG response using Eqs. (14) – (17) with weight coefficients in the form of Eq. (20) seems problematic because the values of the components of second- and third-order susceptibility tensors and Fresnel factors are hardly predictable. However, below we present a procedure that allows analyzing this effect with respect to the bulk parameters. Here we assume that all of the components of second- and third-order susceptibility tensors and Fresnel factors are identical to those of the single crystals of $Bi_2Se_3(111)$ and only DC-electric-field-induced effects are important. Figure 6 shows the thickness dependence of the integral SHG intensity, which has been measured with the fixed incident laser power. The general behavior of the dependence indicates that the narrow resonance feature of SHG enhancement observed for the 10 nm thick film is superimposed on the much broader enhancement trend with decreasing film thickness that extends over the entire range of the film thicknesses. The latter enhancement is about 10-fold with decreasing film thickness in the range from 40 to 6 nm. For the 10 nm thick film, the SHG intensity is resonantly enhanced another ~10-fold. The total enhancement of the SHG response for the 10 nm thick film as compared to the 40 nm thick film, which for many measured parameters roughly corresponds to bulk,[34-36] is about 100-fold. This thickness dependence clearly demonstrates that there are two different DC-electric-field-induced sources contributing to the EFISHG response.

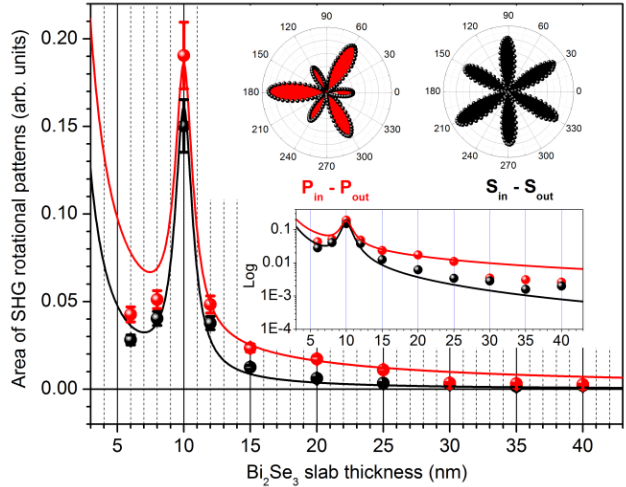

Fig. 6. (Color online) $Bi_2Se_3$ slab thickness ($d$) dependence of the integral SHG intensity (the area of SHG rotational anisotropy patterns) measured with the fixed incident laser power in the $S_{in}$ - $S_{out}$ (black dots) and $P_{in}$ - $P_{out}$ (red dots) light polarization geometries. The solid curves present the best fits to the data using the quadratic and linear+quadratic $\mathcal{E}_{DC}$ field contributions [Eq. (23)]. A Lorentz-shaped resonance peak at 10 nm of ~1.6 nm FWHM is the same amplitude for both fits. Examples of the area of SHG rotational anisotropy patterns measured in the $P_{in}$ - $P_{out}$ and $S_{in}$ - $S_{out}$ polarization geometry for the 10 nm thick film are shown in upper insets by the corresponding colors. The lower inset shows the same dependences but in semi-log scale.

The DC electric field that results from a dynamical redistribution of photoexcited long-lived carriers between the film surfaces seems to be readily applicable for the broad enhancement dynamics.[34,35] Because the photoexcited carrier life-time exceeds the inversed repetition rate of the laser, a quasi-steady Fermi energy difference at the opposite-surface Dirac SS can be reached since the carrier excitation profile in the film follows the Beer's law and therefore the density of photoexcited carrier on the top surface of $Bi_2Se_3$ films is about number-$e$ times larger than that on the bottom surface. The development of the resulting capacitor-type electric field is mediated by the depletion of 3D carrier (both free and photoexcited) when the sum of depletion layer widths associated with each surface of the film exceeds the film thickness. Subsequently, the SHG enhancement with decreasing film thickness can be explained as an EFISHG response associated with the capacitor-type electric field, which persists as long as the sample is irradiated with laser light and is directly related to the film thickness by[35]

$$\mathcal{E}_{DC} = (2E_F - \Delta)/ed, \quad (23)$$

where $E_F$ is the Fermi energy of photoexcited carriers on the top surface of the film, $\Delta$ is the relative shift of Dirac points at the opposite surfaces (equal to the difference in Fermi energies), $e$ is electron charge, $d$ is the film thickness. Consequently, the capacitor-type DC electric field is expected to increase with decreasing film thickness as $\mathcal{E}_{DC} \propto 1/d$.

The resonant feature can also be explained as an EFISHG response but induced by the DC electric field associated with a nonlinearly excited Dirac plasmon.[36] We note that because the phase velocity of 2D plasmons is typically much lower than the velocity of light, the electric field acting on the near-surface region due to the Dirac plasmon excitation can also be considered as a DC electric field. For graphene and semiconductor 2D electron systems, the excitation of a plasmon has been predicted to provide a huge Lorentz-type resonance in SHG intensity at the plasmon frequency.[49] The Dirac plasmon excitation is expected to occur within a four-wave mixing process (stimulated Raman scattering),[36] which is enhanced by a direct optical coupling of incident laser light to Dirac SS at the resonance energy of ~1.5 eV.[64]

This direct optical coupling to Dirac electronic states is expected to initiate a significant vibrational pumping of the Dirac SS through the nonlinear coherent mechanism,[65-67] which is known to occur in continuous-wave (CW) and pulsed Raman lasers.[68] Consequently, the stimulated Raman loss is expected to govern a laser intensity attenuation due to an extra absorption of incident laser light in the free-carrier population at the resonance frequency of a Raman-active surface mode[65] and therefore can be a source of the Dirac plasmon excitation. Because the free-carrier density is found to increase with decreasing film thickness,[36] the Dirac-plasmon-induced SHG effect can be reached for a certain free-carrier density, corresponding to the 10 nm thick film.

We also note that similar resonant features for the 10 nm thick $Bi_2Se_3$ film have been observed in the thickness dependences of the transient reflectivity signal intensities,[34,35] the ultrafast carrier relaxation rates,[34] and the Raman surface phonon mode linewidth.[36] In the latter case the resonant feature has been explained as being resulted from quantum interference between continuum electronic states of a Dirac plasmon and the discrete-energy surface phonon mode. The manifestation of the resonance-type thickness dependence for thin-films of the TI $Bi_2Se_3$ in several independent experimental techniques suggests that a same-origin effect contributes to all of the measured optical responses. However, the most prominent resonance is observed through the EFISHG response, supporting the idea of about the nonlinear origin of the Dirac plasmon excitation since all of the processes involve third-order optical susceptibilities.

To apply the aforementioned theoretical treatment of the rotational anisotropy patterns for thin films of the TI $Bi_2Se_3$, we first use our modeling procedure implemented through Eqs. (14) – (17) with weight coefficients in the form of Eq. (20) to the experimental data on the single crystals of $Bi_2Se_3(111)$.[21] This approach allowed us to obtain the reasonable weight coefficients, which include the surface and depletion-electric-field-induced effects. Subsequently, any changes in weight coefficients with decreasing $Bi_2Se_3$ film thickness in the range from 40 to 6 nm will indicate the finite-size effect on the SHG response intensity, which include EFISHG contributions induced by the capacitor-type and Dirac plasmon-related DC electric fields. Figure 7(a) shows the rotational anisotropy patterns measured for the single crystals of $Bi_2Se_3$(111) (Ref. 21) and the theoretically obtained rotational anisotropy patterns using Eqs. (14) - (17) with the following weight coefficients: $c_1 = 0.0625$, $c_2 = 0.48$, $c_3 = d_3 = 1$, $d_1 = 0.0049$, and $d_2 = 0.16$. Moreover, our theoretical treatment allowed us to obtain the partial contributions of the isotropic, threefold, and sixfold rotational anisotropy components into the total rotational anisotropy patterns measured in the $P_{in} - P_{out}$ and $S_{in} - P_{out}$ light polarization geometries [Fig. 7(b)]. One can see that the isotropic component is very weak, that is in full agreement with rhombohedral nature of $Bi_2Se_3$. The most powerful contribution to the SHG rotational anisotropy patterns is governed by the sixfold component. A much weaker threefold component, being an odd function, simultaneously enhances and suppresses the corresponding rotational peaks of the sixfold component under component summation, giving rise to the dominating threefold rotational symmetry. However, because the threefold components are of opposite signs [Eqs. (16) and (17)], the rotational anisotropy symmetry observed in the $P_{in} - P_{out}$ polarization geometry is rotated by 180° with respect to that observed in the $S_{in} - P_{out}$ polarization geometry. We note also that according to the aforementioned theoretical consideration, the sixfold rotational anisotropy component measured in the $P_{in} - P_{out}$ and $S_{in} - P_{out}$ light polarization geometries is indeed rotated by 30° with respect to the sixfold rotational anisotropy component measured in the $S_{in} - S_{out}$ and $P_{in} - S_{out}$ light polarization geometries [Figs. 7(a) and (b)].

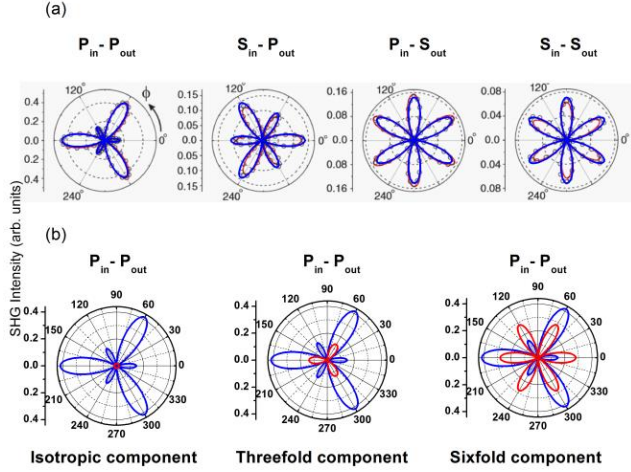

Fig. 7. (Color online) (a) SHG rotational anisotropy patterns measured for different light polarization geometries (open circles) are shown together with theoretical fits (red lines).[21] Theoretical modeling of the SHG rotational anisotropy patterns using Eqs. (14) – (17) with the corresponding weight coefficients of the form of Eq. (20) is shown as blue lines. (b) An example of the theoretically obtained total SHG rotational anisotropy pattern for the $P_{in} - P_{out}$ light polarization geometry [the same as that shown in (a)] (blue) and the corresponding partial contributions of the isotropic, threefold, and sixfold components (red). The sum of components shown in red gives the total SHG intensity shown in blue. Because threefold component is odd, its negative part does not manifest itself in the polar plot, whereas it contributes to the total SHG signal under component summation.

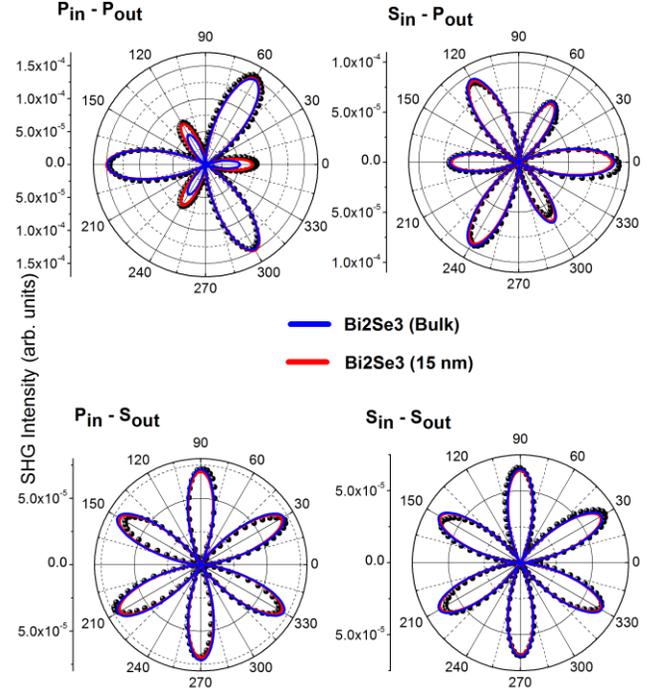

Fig. 8. (Color online) The rotational anisotropy patterns for the 15 nm thick $Bi_2Se_3$ film (black dots) measured for different light polarization geometries indicated. The blue lines present the theoretical traces, which were taken using Eqs. (14) – (17) and the corresponding bulk parameters of $Bi_2Se_3$ [Fig. 7(a)]. The red curves present the theoretical fits for the 15 nm thick film.

Now we apply our fitting procedure to the SHG rotational anisotropy patterns of thin films. Figure 8 shows an example of the fit for the 15 nm thick film, which indicates that the parameters obtained for the single crystals of $Bi_2Se_3$(111) reproduce the rotational anisotropy patterns of the 15 nm thick film, except for the $P_{in} - P_{out}$ polarization geometry where the weight coefficient $c_2 = 0.48$ for the bulk is decreased to $c_2 = 0.36$. The latter behavior implies that the threefold rotational anisotropy component measured in the $P_{in} - P_{out}$ light polarization geometry becomes weaker. Subsequently, because the sixfold component dominates the dynamics [Fig. 7(b)], the resulting shape of the rotational anisotropy pattern with decreasing film thickness becomes more like the sixfold one. We note that these dynamics can be recognized exclusively owing to the fitting procedure used because visually it looks like there is an increase of the intensity of the three weak peaks of the rotational anisotropy pattern when switching from the single crystals of $Bi_2Se_3(111)$ to the 15 nm thick film (Fig. 8).

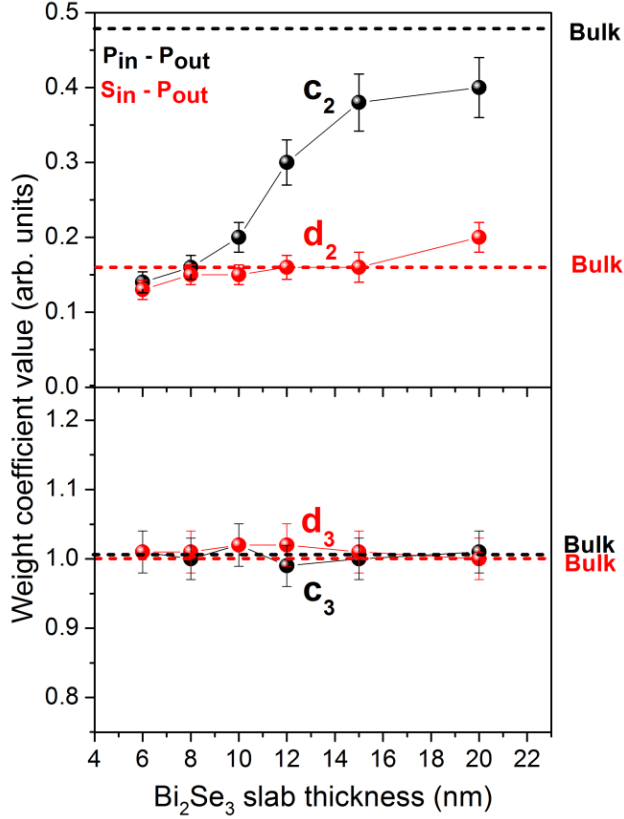

FIG. 9. (Color online) $Bi_2Se_3$ film thickness dependence of weight coefficients (see text) of SHG responses obtained from the fits of the rotational anisotropy patterns measured in the $P_{in}$ - $P_{out}$ and $P_{in}$ - $S_{out}$ light polarization geometries. The same color straight dashed lines indicate the corresponding bulk parameters [Fig. 4(a)].

The similar trends were observed for other samples of various thicknesses. Figure 5 shows the application of the fitting procedure to all of the rotational anisotropy patterns measured. One can see that the aforementioned effect progressively increases with decreasing film thickness, indicating that the threefold rotational anisotropy component measured in the $P_{in}$ - $P_{out}$ light polarization geometry gradually decreases. If the $S_{in}$ - $P_{out}$ light polarization geometry is applied, the effect becomes much weaker. Alternatively, the film thickness change does not affect significantly the shape of the rotational anisotropy patterns measured in the $P_{in}$ - $S_{out}$ and $S_{in}$ - $S_{out}$ light polarization geometries, which always show the sixfold rotational symmetry. The latter behavior implies that the enhancement dynamics measured in $P_{in}$ - $S_{out}$ and $S_{in}$ - $S_{out}$ light polarization geometries is governed by coefficient $b^{(1)}$, which is assumed to be $b^{(1)} \gg a^{(1)}$, and the sign of this coefficient does not affect the weight coefficient $c_3$ since its quadratic form [Eq. (20)]. Similarly, the weight coefficients $c_1$ and $d_1$, which are responsible for isotropic component measured in the $P_{in}$ - $P_{out}$ and $S_{in}$ - $P_{out}$ light polarization geometries, remain always positive and therefore do not affect significantly the shape of the SHG rotational anisotropy patterns.

These dynamics of the rotational anisotropy patterns with decreasing film thickness is demonstrated explicitly in Fig. 9, where the film thickness dependences of the weight coefficients responsible for the threefold and sixfold rotational anisotropy components measured in the $P_{in}$ - $P_{out}$ and $S_{in}$ - $P_{out}$ light polarization geometries are presented. Despite the general SHG enhancement trends, the sixfold rotational anisotropy components (weight coefficients $c_3$ and $d_3$) remain constant with decreasing film thickness (rotational anisotropy patterns shown in Fig. 5 were normalized) and correspond to those of the single crystals of $Bi_2Se_3(111)$. In contrast, the threefold component (weight coefficients $c_2$ and $d_2$) gradually decreases with respect to the bulk parameters. The effect is more significant for the $P_{in}$ - $P_{out}$ light polarization geometry as compared to the $S_{in}$ - $P_{out}$ one (Fig. 9). We note that a similar tendency in the shape change of the SHG rotational anisotropy patterns has been observed for the single crystals of $Bi_2Se_3(111)$ after cleavage as a consequence of DC electric field development due to $O_2$ molecule adsorption at a surface Se vacancy site.[21] This discrepancy between the SHG responses measured in the $P_{in}$ - $P_{out}$ and $S_{in}$ - $P_{out}$ light polarization geometries can be explained by taking into account the fact that both the linear and quadratic DC electric field effects contribute to the EFISHG response. Because $a^{(1)}$ and $b^{(1)}$ coefficients are of opposite signs [Eqs. (13) and (22)], the cross terms in weight coefficients $c_2$ and $d_2$ ($b^{(i)} \otimes b^{(j)}$ and $b^{(i)} \otimes a^{(j)}$) [Eq. (20)] can be either positive or negative. Moreover, the cross term $b^{(i)} \otimes b^{(j)}$ provides a quadratic DC electric field effect, whereas the cross term $b^{(i)} \otimes a^{(j)}$ is responsible for the linear DC electric field effect. If the strength of the DC electric field increases with decreasing film thickness, the weight coefficients $c_2$ and $d_2$ will tend to decrease as a consequence of summations of the opposite-sign cross terms of the different order DC electric field effects. Consequently, the more prominent decrease of weight coefficient $c_2$ with decreasing film thickness as compared to weight coefficient $d_2$ (Fig. 9) indicates that the difference between the positive and negative cross terms in the latter case is smaller [Eq. (20)].

This conclusion is also confirmed by the fit of the thickness dependences of the integral SHG intensity (Fig. 6). Specifically, the integral SHG intensity of the in-plane sixfold rotational anisotropy component measured in the $S_{in}$ - $S_{out}$ light polarization geometry increases quadratically with decreasing film thickness ($\mathcal{E}_{DC}^2 \propto 1/d^2$). This behavior agrees with the theoretical predictions of Eqs. (14), (20), and (22). In this fitting procedure the resonant feature due to the Dirac plasmon excitation has been approximated by a Lorentz-shaped peak of ~1.6 nm FWHM. Keeping the resonance peak of the same intensity and adding only the linear DC electric field term ($\mathcal{E}_{DC} \propto 1/d$), the thickness dependence of the integral SHG intensity measured in the $P_{in}$ - $P_{out}$ light polarization geometry can be acceptably fitted in full agreement with Eqs. (16), (20), and (22).

## IV. CONCLUSIONS

We have provided experimental evidence that the integral intensity of the SHG response from thin films of the TI $Bi_2Se_3$ can be significantly enhanced (about 2 orders of magnitude) with decreasing film thickness from 40 to 10 nm. The enhancement includes the two EFISHG processes: (i) a ~10-fold increase of the SHG intensity with decreasing film thickness from 40 to 6 nm, which is due to the EFISHG contribution induced by the 3D-carrier-depletion-mediated capacitor-type DC electric field, the strength of which gradually increases owing to the dynamical charge imbalance between the opposite-surface Dirac SS and (ii) another ~10-fold resonant enhancement of the EFISHG intensity for the 10 nm thick film with the Lorentz-shaped resonance of ~1.6 nm FWHM, which is due to the DC electric field associated with a nonlinearly excited Dirac plasmon. Because the Dirac plasmon frequency is tunable with the density of free-carrier in the films, which in turn increases with decreasing film thickness, the resonant feature can be observed for a certain free-carrier density corresponding to the 10 nm thick film. We also observed a relative decrease of out-of-plane contribution to the EFISHG response with respect to the in-plane contribution with decreasing film thickness from 40 to 6 nm, which is associated with the joint contributions of the linear and quadratic DC electric field effects to the EFISHG response.

## ACKNOWLEDGMENTS

This work was supported by a Research Challenge Grant from the West Virginia Higher Education Policy Commission (HEPC.dsr.12.29). Some of the work was performed using the West Virginia University Shared Research Facilities.

*Corresponding author: ydglinka@mail.wvu.edu